\documentclass[aps,nofootinbib,prd,superscriptaddress,tightenlines,notitlepage,twocolumn,showpacs,longbibliography]{revtex4-1}

\usepackage{graphicx} 
\usepackage[usenames,dvipsnames]{xcolor}
\usepackage{xspace} 
\usepackage{bm} 
\usepackage{amsmath,amsfonts,amssymb,amsthm}
\usepackage[utf8]{inputenc} 

\xdefinecolor{mylinkcolor}{rgb}{0,0,0.5}
\usepackage[
	bookmarksnumbered, bookmarksopen, bookmarksopenlevel=2,
	breaklinks=true,
	colorlinks=true, filecolor=mylinkcolor, citecolor=mylinkcolor,
	linkcolor=mylinkcolor, urlcolor=mylinkcolor, menucolor=mylinkcolor,
]{hyperref}

\def\vct#1{{\bm{#1}}}

\def\nl{\\ & \quad}

\def\nnlq{\nonumber \\ & \quad \qquad}

\DeclareMathOperator{\Order}{\mathcal{O}}
\DeclareMathOperator{\s}{\mathcal{S}} 
\DeclareMathOperator{\mn}{\mathcal{M}} 

\def\Q{Q} 
\def\q{q} 
\def\e{\mathcal{E}} 
\def\scalar{\phi} 
\def\scals{\phi^\text{IR}} 
\def\scalf{\phi^\text{UV}} 
\def\scal0{\phi_0} 
\def\m{m} 
\def\y{y} 
\def\sig{\tau} 

\allowdisplaybreaks

\newcommand{\AEI}{\affiliation{Max Planck Institute for Gravitational Physics (Albert Einstein Institute), Am M\"uhlenberg 1, Potsdam 14476, Germany}}
\newcommand{\Maryland}{\affiliation{Department of Physics, University of Maryland, College Park, Maryland 20742, USA}}

\begin{document}

\title{Theory-agnostic framework for dynamical scalarization of compact binaries}
\hypersetup{pdftitle={Theory-agnostic framework for dynamical scalarization of compact binaries}}

\author{Mohammed Khalil}\email{mohammed.khalil@aei.mpg.de}\AEI\Maryland
\author{Noah Sennett}\email{noah.sennett@aei.mpg.de}\AEI\Maryland
\author{Jan Steinhoff}\email{jan.steinhoff@aei.mpg.de}\AEI
\author{Alessandra Buonanno}\email{alessandra.buonanno@aei.mpg.de}\AEI\Maryland

\date{\today}

\begin{abstract}
Gravitational wave observations can provide unprecedented insight into the fundamental nature of gravity and allow for novel tests of modifications to General Relativity.
One proposed modification suggests that gravity may undergo a phase transition in the strong-field regime; the detection of such a new phase would constitute a smoking-gun for corrections to General Relativity at the classical level.
Several classes of modified gravity predict the existence of such a transition---known as \textit{spontaneous scalarization}---associated with the spontaneous symmetry breaking of a scalar field near a compact object.
Using a strong-field-agnostic effective-field-theory approach, we show that all theories that exhibit spontaneous scalarization can also manifest \textit{dynamical scalarization}, a phase transition associated with symmetry breaking in a binary system.
We derive an effective point-particle action that provides a simple parametrization describing both phenomena, which establishes a foundation for theory-agnostic searches for scalarization in gravitational-wave observations.
This parametrization can be mapped onto any theory in which scalarization occurs; we demonstrate this point explicitly for binary black holes with a toy model of modified electrodynamics.
\end{abstract}

\maketitle

\section{Introduction}
\label{sec:intro}

Classical gravity described by General Relativity (GR) has passed many experimental tests, from the scale of the Solar System~\cite{Will:2014kxa} and binary pulsars~\cite{Wex:2014nva, Kramer:2016kwa} to the coalescence of binary black holes (BHs) \cite{Abbott:2016blz,TheLIGOScientific:2016src,TheLIGOScientific:2016pea,Abbott:2017vtc,LIGOScientific:2019fpa} and neutron stars (NSs) \cite{Abbott:2018lct}.
Despite its observational success, certain theoretical aspects of GR (e.g., its nonrenormalizability and its prediction of singularities \cite{Hawking:1969sw}) impede progress toward a complete theory of quantum gravity; yet, strong-field modifications of the theory may alleviate these issues \cite{Stelle:1976gc}.

Gravitational wave (GW) observations probe the nonlinear, strong-field behavior of gravity and thus can be used to search for (or constrain) deviations from GR in this regime.
Because detectors are typically dominated by experimental noise, sophisticated methods are required to extract GW signals.
The most sensitive of these techniques rely on modeled predictions of signals (gravitational waveforms), which are matched against the data.
This same approach can be adopted to test gravity with GWs; to do so requires accurate signal models that faithfully incorporate the effects from the strong-field deviations one hopes to constrain~\cite{Will:2014kxa}.
Ideally, these models would be agnostic about details of the strong-field modifications to GR, so that a single test could constrain a variety of alternative theories of gravity.

This work establishes a framework for such tests given the hypothetical scenario in which the gravitational sector manifests phase transitions, with only one phase corresponding to classical GR.
This proposal comprises an attractive target for binary pulsar and GW tests of gravity; if the transition between phases arises only in the strong-gravity regime (e.g., in the presence of large curvature, relativistic matter, etc.), then such a theory could generate deviations from GR in compact binary systems while simultaneously evading stringent constraints set by weak-gravity tests.
We consider the case wherein the ``new'' phases arise via spontaneous symmetry breaking in the gravitational sector.
Similar phase transitions occur in many areas of contemporary physics---perhaps the most famous example is the electroweak symmetry breaking through the Higgs field~\cite{Englert:1964et,Higgs:1964pj,Guralnik:1964eu}---so it is sensible to consider their appearance in gravity as well.
As a first step, we focus on a simple set of such gravitational theories, in which the transition from GR to a new phase most closely resembles the spontaneous magnetization of a ferromagnet; however, these theories can also be extended to instead replicate the standard Higgs mechanism in the gravitational sector \cite{Coates:2016ktu,Ramazanoglu:2018tig}.

Specifically, we investigate the nonlinear \textit{scalarization} of nonrotating compact objects (BHs and NSs), which arises from spontaneous symmetry breaking of an additional scalar component of gravity~\cite{Damour:1996ke,Sennett:2017lcx}.
\textit{Spontaneous scalarization}---the scalarization of a single, isolated object---has been found in several scalar extensions of GR, including massless \cite{Damour:1993hw,Stefanov:2007eq,Doneva:2010ke,Palenzuela:2015ima,Mendes:2016fby,Herdeiro:2018wub,Fernandes:2019rez} and massive \cite{Chen:2015zmx,Ramazanoglu:2016kul,Sagunski:2017nzb} scalar-tensor (ST) theories and extended scalar-tensor-Gauss-Bonnet (ESTGB) theories \cite{Doneva:2017bvd,Silva:2017uqg,Antoniou:2017acq,Antoniou:2017hxj,Doneva:2017duq,Brihaye:2019kvj}.
Similar phenomena can also occur for vector \cite{Ramazanoglu:2017xbl, Ramazanoglu:2019gbz}, gauge \cite{Ramazanoglu:2018tig}, and spinor \cite{Ramazanoglu:2018hwk} fields.
In contrast, \textit{dynamical scalarization}---scalarization that occurs during the coalescence of a binary system---has been demonstrated and modeled only for NS binaries in ST theories \cite{Barausse:2012da,Palenzuela:2013hsa,Shibata:2013pra,Sampson:2014qqa,Taniguchi:2014fqa,Zimmerman:2015hua,Sennett:2016rwa,Sennett:2017lcx}.

A scalarized compact object emits scalar radiation when accelerated, analogous to an accelerated electric charge.
In a binary system, the emission of scalar waves augments the energy dissipation through the (tensor) GWs found in GR, hastening the orbital decay.
Radio observations of binary pulsars \cite{Wex:2014nva, Kramer:2016kwa,Shao:2017gwu,Anderson:2019eay} and GW observations of coalescing BHs and NSs \cite{Abbott:2018lct,LIGOScientific:2019fpa} are sensitive to anomalous energy fluxes, and thus can be used to constrain the presence of scalarization in such binaries.
Binaries containing spontaneously scalarized components emit scalar radiation throughout their entire evolution.
In contrast, dynamically scalarizing binaries transition from an unscalarized (GR) state to a scalarized (non-GR) state at some critical orbital separation, only emitting scalar waves after this point.
Because this is a second-order phase transition~\cite{Sennett:2017lcx}, the emitted GWs contain a sharp feature corresponding to the onset of dynamical scalarization.
This feature cannot be replicated within typical theory-agnostic frameworks used to test gravity~\cite{Arun:2006hn,Yunes:2009ke, Li:2011cg, Agathos:2013upa}, as these only consider smooth deviations from GR predictions, e.g., modifications to the coefficients of a power-series expansion of the phase evolution.

While one could attempt to model dynamical scalarization phenomenologically by adding nonanalytic functions to such frameworks \cite{Sampson:2013jpa,Sampson:2014qqa}, in this paper, we propose a complementary theory-agnostic approach.
We focus on a specific non-GR effect, here scalarization, but remain agnostic toward the particular alternative theory of gravity in which it occurs.
The basis for our framework is effective field theory.
Scalarization arises from strong-field, nonlinear scalar interactions in the vicinity of compact objects; the details of this short-distance physics depends on the specific alternative to GR that one considers.
By integrating out these short-distance scales, we construct an effective point-particle action for scalarizing bodies in which the relevant details of the modification to GR are encapsulated in a small set of form factors.
The coefficients of these couplings offer a concise parametrization ideal for searches for scalarization with GWs.
The essential step in constructing this effective theory is identifying the fields and symmetries relevant to this phenomenon.
Starting from the perspective that scalarization coincides with the appearance of a tachyonic scalar mode of the compact object, we derive the unique leading-order effective action valid near the critical point of the phase transition.
Though this effective action matches that of Ref.~\cite{Sennett:2017lcx}---which describes the scalarization of NSs in ST theories\footnote{Dynamical scalarization was also modeled at the level of equations of motion in Refs.~\cite{Palenzuela:2013hsa, Sennett:2016rwa}.}---the approach described here is valid for a broader range of non-GR theories.

Our proposed parametrization of scalarization is directly analogous to the standard treatment of tidal interactions in compact binary systems.
Tidal effects enter GW observables through a set of parameters that characterize the response of each compact object to external tidal fields~\cite{Flanagan:2007ix}.
These parameters are determined by the structure of the compact bodies---for example, the short-distance nuclear interactions occurring in the interior of a NS.
This description of tidal effects is applicable to a broad range of nuclear models (i.e., NS equations of state) and offers a more convenient parametrization of unknown nuclear physics for GW measurements~\cite{TheLIGOScientific:2017qsa,Abbott:2018wiz} than directly incorporating nuclear physics into GW models.
From the perspective of modeling compact binaries, the primary difference between tidal effects and scalarization is that the latter is an inherently nonlinear phenomenon, necessitating higher-order interactions in an effective action.

Beyond offering a convenient parametrization for GW tests of gravity, our effective action also elucidates certain generic properties of scalarization phenomena.
Using a simple analysis of energetics based on the effective theory, we argue that \emph{any theory that admits spontaneous scalarization must also admit dynamical scalarization}.
Additionally, this type of analysis can provide further insights regarding the (nonperturbative) stability of scalarized configurations and the critical phenomena close to the scalarization phase transition.
We illustrate these points by applying our energetics analysis to a simple Einstein-Maxwell-scalar (EMS) theory in which electrically charged BHs can spontaneously scalarize, complementing previous results for NSs in ST theories \cite{Sennett:2017lcx}.

The paper is organized as follows.
In Sec.~\ref{sec:effaction}, we first review the mechanism of scalarization as the spontaneous breaking of the $\mathbb{Z}_2$ symmetry of a scalar field driven by a linear scalar-mode instability.
Then, we construct an effective worldline action for a compact object interacting with a scalar field valid near the onset of scalarization.
In Sec.~\ref{sec:matching}, we discuss how the relevant coefficients in the action can be matched to the energetics of an isolated static compact object in an external scalar field, demonstrating the procedure explicitly with BHs in the EMS theory of Ref.~\cite{Herdeiro:2018wub}.
In Sec.~\ref{sec:strongeffects}, we employ the effective action to further investigate scalarization in this EMS theory: we examine the stability of scalarized configurations, compute the critical exponents of the scalarization phase transition, and predict the frequency at which dynamical scalarization occurs for binary BHs.
We also argue that dynamical scalarization is as ubiquitous as spontaneous scalarization in modified theories of gravity.
Finally, in Sec.~\ref{sec:conclusions}, we summarize the main implications of our findings, and discuss future applications of our framework.
The appendixes provide a derivation of a more general effective action and details on the construction of numerical solutions for isolated BHs in EMS theory.\footnote{Throughout this work, we use the conventions of
  Misner, Thorne, and Wheeler~\cite{Misner:1974qy} for the metric
  signature and Riemann tensor and work in units in which the speed of
  light and bare gravitational constant are unity.}

\section{Linear mode instability and effective action close to critical point}
\label{sec:effaction}

In this section, we review the connection between the appearance of an unstable scalar mode in an unscalarized compact object and the existence of a scalarized state for the same body. We then derive an effective action close to this critical point at which this mode becomes unstable.

As an illustrative toy model for this discussion, we consider the modified theory of electrodynamics introduced in Ref.~\cite{Herdeiro:2018wub} (hereafter referred to as EMS theory for brevity), whose action is given by
\begin{equation} \label{EMSaction}
  S_\text{field} = \int d^4x \, \frac{\sqrt{-g}}{16\pi} \left[ R - 2 \partial_\mu \scalar \partial^\mu \scalar - f(\scalar) F^{\mu\nu} F_{\mu\nu} \right],
\end{equation}
where $R$ is the Ricci scalar, $g$ is the determinant of the metric $g_{\mu\nu}$, and $F_{\mu\nu} = \partial_\mu A_\nu - \partial_\nu A_\mu$ is the electromagnetic field tensor.
In this paper, we consider two choices of scalar couplings:
\begin{align}
f_1(\phi) =& e^{-\alpha\scalar^2}, \label{eq:f1coupling} \\
f_2(\phi) = &\left(1+\alpha \scalar^2 -\frac{1}{2} \alpha^2 \phi^4\right)^{-1}, \label{eq:f2coupling}
\end{align}
where $\alpha$ is a dimensionless coupling constant.
While the two couplings have the same behavior near $\phi =0$, their behavior for large field value differs drastically.
The former choice was used in Ref.~\cite{Herdeiro:2018wub} to construct stable scalarized BH solutions, whereas we introduce the latter in this work to demonstrate a theory in which no stable scalarized BH configurations exist (see Sec.~\ref{sec:strongeffects}).

The absence of any linear coupling of $\phi$ to the Maxwell term implies that any solution in Einstein-Maxwell (EM) theory, i.e., with $\scalar = 0$, also solves the field equations of Eq.~\eqref{EMSaction}; however, stable solutions in EM theory may be unstable in EMS theory.
To see this, we write the scalar-field equation schematically as
\begin{equation}
  \Box \scalar = m_\text{eff}^2 \, \scalar , \qquad \label{eq:EffMass}
  m_\text{eff}^2 = \frac{f'(\phi)}{4 \phi} F^{\mu\nu} F_{\mu\nu} .
\end{equation}
We consider an electrically charged BH, for which ${F^{\mu\nu} F_{\mu\nu} < 0}$ and thus the effective-mass squared $m_\text{eff}^2$ is negative for $\alpha <0$. 
One can decompose $\scalar$ into Fourier modes with frequency $\omega$ and wave vector $\vct{k}$, which satisfy the dispersion relation $\omega^2 \approx \vct{k}^2 + m_\text{eff}^2(\vct{k})$, where curvature corrections have been dropped for simplicity.
We see that if $m_\text{eff}^2(\vct{k})$ is sufficiently negative, then $\omega^2$ is also negative, leading to a tachyonic instability.
The critical point at which this tachyonic instability first appears can be determined by identifying linearly unstable quasinormal scalar modes of the EM solution~\cite{Silva:2017uqg,Myung:2018vug} or by constructing sequences of fully nonlinear, static scalarized solutions (as we do here)~\cite{Herdeiro:2018wub,Myung:2018vug}.

The tachyonic instability drives the body away from the unscalarized solution, thereby breaking the symmetry $\scalar \rightarrow - \scalar$ in Eq.~\eqref{EMSaction}.
For a stable scalarized equilibrium configuration to exist, this instability must saturate in the nonlinear regime \cite{Ramazanoglu:2017yun}.
These two conditions---the existence of a tachyonic instability and its eventual saturation---are satisfied in all of the theories discussed previously~\cite{Damour:1993hw,Stefanov:2007eq,Doneva:2010ke,Palenzuela:2015ima,Mendes:2016fby,Chen:2015zmx,Ramazanoglu:2016kul,Sagunski:2017nzb,Doneva:2017bvd,Silva:2017uqg,Antoniou:2017acq,Antoniou:2017hxj,Doneva:2017duq,Brihaye:2019kvj,Herdeiro:2018wub,Fernandes:2019rez,Ramazanoglu:2017xbl,Ramazanoglu:2019gbz,Ramazanoglu:2018tig,Ramazanoglu:2018hwk}. 
The only difference between these theories is the form of $m_\text{eff}^2$; for example, in ST theories $m_\text{eff}^2$ depends on the stress-energy tensor, while in ESTGB it depends on the Gauss-Bonnet invariant.
Indeed, theories which meet these two criteria can be straightforwardly constructed, which is the reason why scalarization is such a ubiquitous phenomenon.

\begin{figure*}
\includegraphics{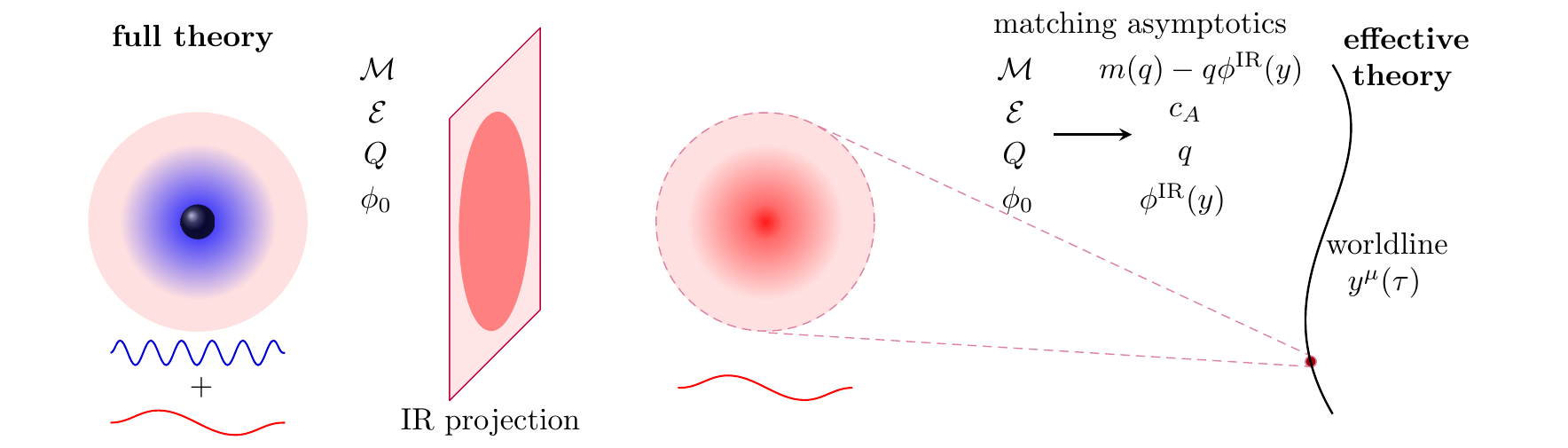}
\caption{\label{fig:matching} An illustration of the approach used in this paper (read from left to right).
  (\emph{Left}) The fields describing an isolated compact object in equilibrium are decomposed into short- (UV) and long- (IR) wavelength modes, depicted in blue and red, respectively.
  (\emph{Center}) We integrate out the UV modes via an IR projection.
  (\emph{Right}) By matching asymptotics, we identify the coarse-grained compact object with an effective point-particle model that describes the IR sector of the full theory.
  Section~\ref{sec:matching} contains a detailed description of this procedure and the definitions of all quantities shown above.
}
\end{figure*}

An even simpler perspective on scalarization arises from a coarse-grained, or effective, theory. Let us derive it explicitly.
We start by splitting the fields into the short- (or ultraviolet, UV) and long- (or infrared, IR) wavelength regimes separated by the object's size $\sim R$, i.e., $\scalar = \scals + \scalf$, and spatially average over (integrate out) the UV parts.
This effectively shrinks the compact object to a point and its effective action is given by an integral over a worldline $\y^\mu(\sig)$, where $\sig$ is the proper time (see Fig.~\ref{fig:matching} for a schematic illustration).
Dynamical short-length-scale processes like oscillations of the object are represented by dynamical variables on the worldline.
For simplicity, we assume that we can also average over fast oscillation modes and only retain the monopolar mode associated to a linear tachyonic instability, denoted by $\q(\sig)$.
This mode $\q(\sig)$ can indeed be excited by IR fields, since its frequency (or effective mass) vanishes at the critical point.

Effective actions are usually constructed by making an ansatz respecting certain symmetries and including only terms up to a given power in the cutoff between IR and UV scales.
The relevant symmetries here are diffeomorphism, U(1)-gauge, worldline-reparametrization, time-reversal\footnote{Time reversal is an approximate symmetry of compact objects in an adiabatic setup, like the inspiral of a binary system. In this case, a compact object's entropy remains approximately constant.}, and scalar-inversion invariance.
The last reads $\scalar \rightarrow -\scalar$ in the full theory, so in the effective action it decomposes into simultaneous IR $\scals \rightarrow - \scals$ and UV $\q \rightarrow -\q$ transformations.
The IR fields are of order $\scals \sim \Order(R/r)$ on the worldline, where $r$ is the typical IR scale (e.g., the separation of a binary).
Now, the oscillator equation for the mode $q(\sig)$ driven by the IR field $\scals$ can be schematically written as
\begin{equation}
  c_{\dot{\q}^2} \ddot{\q} + V'(q) = \scals(y) , \quad
  V(q) = \frac{c_{(2)}}{2} q^2 + \frac{c_{(4)}}{4!} q^4 + \dots ,\label{eq:effEOM}
\end{equation}
where $\dot{~} = d / d \sig$ and the $c_{\dots}$ are constant coefficients determined by the UV physics. 
(The singular self-field contribution to $\scals(y)$ must be removed using some regularization prescription.)
The normalization of $\q$ is chosen to fix the coefficient of $\scals(y)$; for all that follows, we simply assume that $c_{\dot{\q}^2} >0$.
Close to the critical point, the quadratic term in $V$ is negligible, and thus from Eq.~\eqref{eq:effEOM}, one finds that for equilibrium configurations ($\dot{\q} =0$), $\q$ scales as $\q^3 \sim \scals$.
More generally, the mode $\q$ oscillates around this equilibrium point provided that the IR field evolves slowly relative to the frequency of the mode, i.e., $\dot{\phi}^\text{IR} = \dot{y}^\mu \partial_\mu \scals \ll \dot{\q} / \q$; this condition is satisfied for binary systems on quasicircular orbits (which we restrict our attention to in this work), but could be violated for highly eccentric orbits.
For small perturbations around equilibrium, one finds that $\dot{\q} \sim \delta \sqrt{\scals}$ and $\ddot{\q} \sim \delta \scals$ where $\delta \equiv (\q-\q_0)/\q_0 \ll 1$ is the fractional deviation from the equilibrium point $\q_0$.

Using the scaling relations derived above, we construct the most generic effective action for a nonrotating compact object with a dynamical mode $\q(\sig)$ described by Eq.~\eqref{eq:effEOM} close to the critical point, up to order $ \Order(R^2/r^2)$
\begin{align}
  S_\text{CO}^\text{crit} &= \int d\sig \bigg[
  \frac{c_{\dot{\q}^2}}{2} \dot{\q}^2 + \scals(y) \q - c_{(0)}
  - \frac{c_{(2)}}{2} \q^2 - \frac{c_{(4)}}{4!} \q^4 \nnlq
  + c_A A_\mu^\text{IR}(\y) \dot{\y}^\mu
  + \Order\bigg(\frac{R^2}{r^2}\bigg) \bigg]\,, \\
  &= \int d\sig \bigg[
  \frac{c_{\dot{\q}^2}}{2} \dot{\q}^2 + \scals(y) \q - m(q) \nnlq
  + c_A A_\mu^\text{IR}(\y) \dot{\y}^\mu
  + \Order\bigg(\frac{R^2}{r^2}\bigg) \bigg]\,, \label{SCO}
\end{align}
where CO stands for compact object and for later convenience we define
\begin{equation}
  m(q) \equiv c_{(0)} + V(\q) .
\end{equation}
Terms containing time derivatives of $\scals$ all enter at higher order in $R/r$ than we work, e.g., $\ddot{\phi}^\text{IR} \q \sim \dot{\phi}^\text{IR} \dot{\q} \sim \scals \ddot{\q} \sim \Order(R^2 / r^2)$, and thus are absent in Eq.~\eqref{SCO}.
A reparametrization-invariant action is obtained by inserting $d\sig = d \sigma \sqrt{ - g_{\mu\nu}^\text{IR}(\y) d \y^\mu/ d \sigma d\y^\nu / d \sigma}$, where $\sigma$ is an arbitrary affine parameter, and replacing derivatives $d/d\tau$ accordingly.
The complete effective action reads
\begin{align}
  S_\text{eff} = S_\text{field}^\text{IR} + S_\text{CO}^\text{crit}, \label{EFTaction}
\end{align}
where more copies of $S_\text{CO}^\text{crit}$ can be added depending on the number of objects in the system and $S_\text{field}^\text{IR}$ is given by Eq.~\eqref{EMSaction} with IR labels on the fields.
The equations of motion and field equations are obtained by independent variations of $y^\mu(\sigma)$, $q(\sigma)$, and $\scals(x)$, $g_{\mu\nu}^\text{IR}(x)$, $A_\mu^\text{IR}(x)$.

The simplicity of $S_\text{CO}^\text{crit}$ is striking, but we recall that it is only valid close to the critical point of a monopolar, tachyonic, linear instability of a scalar mode.
(A more generic effective action valid away from the critical point is discussed in Appendix \ref{sec:appendixA}.)
Despite its simplicity, the effective action (\ref{EFTaction}) is theory agnostic, in the sense that it is constructed assuming only the scalar-inversion symmetry and that the nonrotating compact object hosts such a mode; in particular, it should hold for the cases studied in Refs.~\cite{Damour:1993hw,Stefanov:2007eq,Doneva:2010ke,Palenzuela:2015ima,Mendes:2016fby,Chen:2015zmx,Ramazanoglu:2016kul,Sagunski:2017nzb,Doneva:2017bvd,Silva:2017uqg,Antoniou:2017acq,Antoniou:2017hxj,Doneva:2017duq,Brihaye:2019kvj,Herdeiro:2018wub,Fernandes:2019rez,Ramazanoglu:2017xbl,Ramazanoglu:2019gbz,Ramazanoglu:2018tig,Ramazanoglu:2018hwk} and similar work to come.
We emphasize that strong-field UV physics at the body scale is parametrized through the numerical coefficients $c_{\dots}$, which can be matched to a specific theory and compact object, or be constrained directly from observations (analogous to tidal parameters \cite{Abbott:2018lct,Abbott:2018wiz}).

\section{Matching strong-field physics into black-hole solutions}
\label{sec:matching}
As an illustrative example of the effective-action framework derived above, we now compute the sought-after coefficients $c_{\dots}$ for BHs in EMS theory.

For this purpose, we match a BH solution in the full theory~\eqref{EMSaction} to a generic solution of the coarse-grained effective theory~\eqref{EFTaction} for an isolated body.
Schematically, the former represents the full solution at all scales, while the latter only represents its projection onto IR scales. We focus first on BH solutions of the full theory~\eqref{EMSaction}, restricting our attention to equilibrium/static, electrically charged, spherically symmetric solutions.

In EMS theory, this family of solutions is characterized by three independent parameters, which we take to be the electric charge $\e$, the BH entropy $\s$, and the asymptotic scalar field $\scal0$, assuming a vanishing asymptotic electromagnetic field and an asymptotically flat metric.
The electric charge is globally conserved by the U(1) symmetry of the theory and the entropy remains constant under reversible processes, which we have implicitly restricted ourselves to by assuming time-reversal symmetry in the effective action.
Thus, we use a sequence of solutions with fixed $\e$ and $\s$ to represent the response of a BH to a varying scalar background $\scal0$.
Since EMS theory modifies electrodynamics, but not gravity or the coupling to gravity, the entropy of a charged BH is the same as in GR, i.e., it is proportional to the horizon area.
See also Ref.~\cite{Herdeiro:2018wub} for the first law of BH thermodynamics in EMS theory.

The asymptotic behavior of the fields take the form
\begin{equation}
  \left(\begin{array}{c}
     \scalar \\ A_0 \\ g_{00}
  \end{array}\right)
  =
  \left(\begin{array}{c}
  \scal0 \\ 0 \\ -1
  \end{array}\right)
  + \left(\begin{array}{c}
  \Q(\phi_0) \\
  - \e e^{\alpha \scal0^2} \\
  2 \mn(\phi_0)
  \end{array}\right) \frac{1}{|\vct{X}|}
  + \Order(|\vct{X}|^{-2}) , \label{eq:UVsols}
\end{equation}
where $\Q(\scal0)$ is the scalar charge of the BH and $\mn(\scal0)$ is its gravitational mass.\footnote{The quantities $\Q(\scal0)$ and $\mn(\scal0)$ describing the asymptotic behavior of the solution also depend on the parameters $\e$ and $\s$, but we suppress the dependence in our notation for brevity. Derivatives of $\Q$ and $\mn$ are taken holding $\e$ and $\s$ constant.}
We construct these solutions numerically (see Appendix~\ref{sec:appendixB} for details), and then compute $\scal0, \mn,$ and $ \Q$ directly from their asymptotic behavior.

\begin{figure*}
  \includegraphics[width=\linewidth]{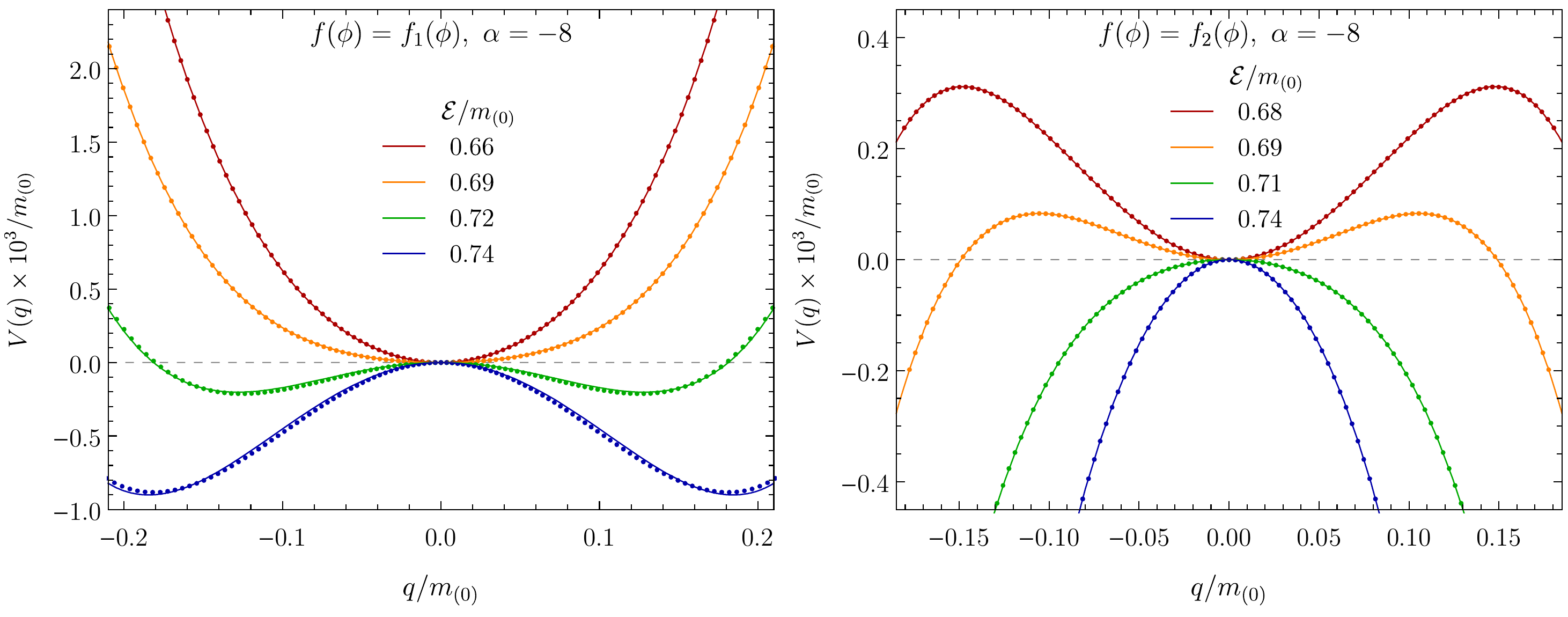}
  \caption{\label{fig:sombrero}The potential $V(\q)$ for $\alpha=-8$ and different electric charges $\e$. The left and right panels correspond to BHs in EMS theory with coupling $f_1(\phi)$~[Eq.~\eqref{eq:f1coupling}] and $f_2(\phi)$~[Eq.~\eqref{eq:f2coupling}], respectively. The numerical solutions and polynomial fits are indicated with points and lines, respectively.}
\end{figure*}

Next, we turn our attention to the description of these BH solutions in the effective theory~\eqref{EFTaction}.
We set up the solution under the same boundary conditions as the numerical sequence described above; in particular, we look at isolated equilibrium configurations, $y^\alpha \partial_\alpha g^\text{IR}_{\mu \nu} \approx y^\alpha \partial_\alpha A_0^\text{IR} \approx y^\alpha \partial_\alpha \scals \approx \dot{\q} \approx 0$.
We construct a coordinate system $\vct{x}$ in which the worldline has spatial components $\vct{y}=\vct{0}$, so that all fields are independent of time.
Furthermore, relying on the fact that both solutions are asymptotically flat, we choose the coordinates $\vct{x}$ such that they match the numerical coordinates $\vct{X}$ in the asymptotic region, i.e., $\vct{x} = \vct{X} + \Order(|\vct{X|}^{-1})$.
Then working to linear order in the fields, we find
\begin{equation}    
  \left(\begin{array}{c}
     \scals \\ A_0^\text{IR} \\ g_{00}^\text{IR}
  \end{array}\right)
  =
  \left(\begin{array}{c}
  \scal0 \\ 0 \\ -1
  \end{array}\right)
  + \left(\begin{array}{c}
  q \\
  - c_A e^{\alpha [\scals(y)]^2} \\
  2 [ m(\q) - \scals(y) \q ]
  \end{array}\right) \frac{1}{|\vct{x}|}
  + \dots . \label{eq:IRsols}
\end{equation}
These fields are singular when evaluated on the worldline, $\vct{x} = \vct{y} = \vct{0}$.
This can be cured by appropriately regularizing the solution; here, we simply keep the finite part and drop the singular self-field part, e.g., $\scals(y) = \phi_0$.
The situation is analogous to the singular fields that arise in electrostatics when an extended source is approximated by a point charge.

In addition to a solution for the fields, a variation of $q$ in the effective action leads to
\begin{equation}
  \scal0 = \frac{d m}{d \q} = \frac{d V}{d \q} = c_{(2)} \q + \frac{c_{(4)} \q^3}{3!} + \Order\bigg(\frac{R^2}{r^2}\bigg) . \label{eq:qEOM}
\end{equation}

The matching now consists of identifying the IR-scale fields in the solution of the full theory~\eqref{eq:UVsols} with the fields predicted in the IR effective theory~\eqref{eq:IRsols}.
We extract the IR-scale fields from the former solution using an appropriate IR projector $P^\text{IR}[\cdot]$, such that the matching conditions are given explicitly as $P^\text{IR}[\scalar] = \scals$ (and likewise for the other fields).
Such a projector is most easily formulated in the Fourier domain, so we first compute the (spatial) Fourier transform of the fields~\eqref{eq:UVsols}, denoted by a tilde
\begin{equation}
  \tilde{\scalar}(\vct{K}) = \scal0 \delta(\vct{K}) + \frac{4\pi Q(\scal0)}{\vct{K}^2} + \Order(|\vct{K}|^{-1}) . \label{eq:phiTilde}
\end{equation}
We employ the simple projector $P^\text{IR}[\tilde{\scalar}] \equiv \tilde{\scalar}(\vct{K}) \Theta(K^\text{IR} - |\vct{K}|)$, where $\Theta$ is the Heaviside function and $K^\text{IR}$ is the cutoff scale.
Applying this projection to Eq.~\eqref{eq:phiTilde} and taking the inverse Fourier transform, one finds that $P^\text{IR}[\scalar]$ takes the same form as Eq.~\eqref{eq:UVsols} on scales longer than the cutoff, i.e., for $|\vct{X}|   \gg 1/ K^\text{IR}$.
Then, our matching conditions $P^\text{IR}[\scalar] = \scals,P^\text{IR}[A_0] = A_0^\text{IR}, P^\text{IR}[g_{00}] = g_{00}^\text{IR} $ reduce to
\begin{equation}
  \Q(\scal0) = \q, \quad
  \e = c_A, \quad
  \mn(\scal0) = m(\q) - \scal0 \q, \label{eq:match}
\end{equation}
where we have used $\scals(y) = \scal0$ as discussed above.
Note that the last equation and $\m'(q) = \scal0$~\eqref{eq:qEOM} reveal that the two measures of energy $\mn$ and $\m$ are related by a Legendre transformation of the conjugate variables $(\q,\scal0)$.
Hence, we find that $\Q = \q = -\mn'(\scal0)$, in agreement with the first law of BH thermodynamics \cite{Gibbons:1996af, Cardenas:2017chu}.

While $\mn(\scal0)$ is the gravitational mass of the system, $\m(\q)$ represents the ``gravitational free energy'' of the body (see also Sec.~III.A of Ref.~\cite{Sennett:2017lcx}).
That is, $\m$ is the mass/energy with the potential energy $- \scal0 \q$ (due to the external scalar field) subtracted from $\mn$.
We find below that $\m$---not $\mathcal{M}$---serves as better representation of ``point-particle mass'' found in the Lagrangian or Hamiltonian description of a binary system; of course, both quantities reduce to the standard ADM mass in GR in the absence of scalarization.
Furthermore, away from the critical point, it is not necessary to treat the mode $q$ as a dynamical variable.
This means that we can set $\dot{\q} = 0$ and remove $\q$ from the action \eqref{SCO}.
The latter is achieved by virtue of the Legendre transformation between $m(\q)$ in Eq.~\eqref{SCO} and $\mn(\scals(y))$,
\begin{multline}\label{eardley}
  S_\text{CO} = \int d\sig \bigg[-\mn(\scals(y))
  + \e A_\mu^\text{IR}(\y) \dot{\y}^\mu + \dots \bigg].
\end{multline}
We see that $\mn$ plays the role of the Eardley mass \cite{eardley1975observable} in the action now.
We note that since the Eardley mass and $\m(q)$ are related by a Legendre transformation, they contain the same information.

To compute the values of the various $c_{\dots}$, we numerically construct a sequence of BH solutions as described above and extract the functions $\mn(\scal0)$ and $\Q(\scal0)$.
From there we obtain $\m(\q)$ and $V(\q)$ numerically from Eq.~\eqref{eq:match}, as illustrated in Fig.~\ref{fig:sombrero}.
Each curve indicates a BH sequence with a different electric charge-to-mass ratio $\e / m^{(0)}$, where $m^{(0)} \equiv c_{(0)}$ is the mass of the isolated BH with no scalar charge.
The points indicate the numerically computed solutions, which are calculated by solving the field equations with different boundary conditions for the scalar field  (see Appendix~\ref{sec:appendixB}).
The solid lines are polynomial fits of the form \eqref{eq:effEOM}, from which we extract the values of the coefficients $c_{(2)}$ and $c_{(4)}$. 

It is remarkable that from equilibrium solutions one can fix the potential of a dynamical (nonequilibrium) mode to order $\q^4$.
This connection is nontrivial, and it breaks down when one relaxes the assumption of being close to the critical point.
For instance, if terms like $(\scals)^2$ are included in Eq.~(\ref{SCO}), then $\Q \neq q$; or consider the case of a minimally coupled scalar field ($\alpha = 0$),  wherein no-hair theorems \cite{Hawking:1972qk,Sotiriou:2011dz} guarantee that $\mn(\scal0) = \text{const}$---the energy of an equilibrium BH obviously does not encode any information about dynamical modes.

Having described how to compute the coefficients $c_{\ldots}$ in the effective action~\eqref{SCO}, the following section illustrates how this action can be used to study spontaneous and dynamical scalarization.

\section{Modeling strong-gravity effects within the theory-agnostic framework}
\label{sec:strongeffects}

In this section, we show how spontaneous and dynamical scalarization can be understood from the effective theory~\eqref{SCO}, based on a simple analysis of energetics.
We use this effective action to investigate the properties of these critical phenomena, namely their critical exponents.
Though we use EMS theory to make quantitative predictions throughout this section, we emphasize again that the qualitative behavior we find should hold generically for all theories in which spontaneous scalarization occurs.
More specifically, theories in which scalarized configurations are stable are analogous to EMS theory with scalar coupling $f_1(\phi)$, whereas those where such configurations are unstable correspond to the coupling $f_2(\phi)$ (see the following subsection for details).
Extending the predictions made below to other theories only requires computation of the effective mode potential $V(q)$, as described in the previous section, and the inclusion of any new long-range fields not present in EMS theory that impact the motion of binary systems.

\subsection{Spontaneous scalarization}

Recall that a spontaneously scalarized object is one that hosts a nonzero scalar charge even in the absence of an external scalar field $\phi_0=0$.
From Eq.~(\ref{eq:qEOM}) we see that $\scal0 = 0$ corresponds to extrema of $V(\q)$ for equilibrium configurations.
Furthermore, since $V(\q)$ is the oscillation-mode potential, $\q$ dynamically evolves into a minimum of $V(\q)$ [see Eq.~\eqref{eq:effEOM}].
Thus, the existence of spontaneously scalarized configurations is indicated by nontrivial extrema of $V(q)$, and the stability of these configurations depends on whether such points are local minima (stable) or maxima (unstable).
For example, the left panel Fig.~\ref{fig:sombrero} depicts the appearance of spontaneously scalarized BH solutions as one increases the charge-to-mass ratio in EMS theory with coupling $f_1(\phi)$.
Without enough electric charge (e.g., the red and orange curves, with $c_{(2)}>0$), the EM (unscalarized) solutions are the only stable BH solutions, but by increasing the charge beyond a critical value (e.g., the green and blue curves, with $c_{(2)}<0$), the EM solution becomes unstable and the stable solutions instead occur at nonvanishing values of $\q$.

Our approach allows one to determine the values of the coupling $\alpha$ and the electric charge $\e$ at which spontaneous scalarization first occurs ($c_{(2)} = 0$) using only sequences of equilibrium BH solutions.
A more direct approach employed in the past was to search for instabilities of linear, dynamical scalar perturbations on a (GR) Reissner-Nordstr\"{o}m background~\cite{Myung:2018vug}.
We find that the two methods provide the same predictions.
For the choice of coupling $f_1(\phi)$, we compute the critical coupling as a function of the electric charge $\alpha_\text{crit}(\e)$ where $c_{(2)} = 0$ and find that our results agree with the predictions of Ref.~\cite{Myung:2018vug} at the onset of the linear instability of the $\ell=0$ scalar mode to within $1\%$.
For theories in which scalarized solutions are easy to construct, like the EMS theory considered here, our approach can more efficiently compute this critical point than a perturbative stability analysis.
We see that the effective potential $V(\q)$ provides strong indications for a linear scalar-mode instability and its nonlinear saturation.

The same energetics argument reveals drastically different behavior in the EMS theory with coupling $f_2(\scalar)$, depicted in the right panel of Fig.~\ref{fig:sombrero}.
Recall that $f_1(\scalar)\approx f_2(\scalar)$ for small field values, but the two choices differ in the nonlinear regime, which will dramatically impact the stability of scalarized solutions.
This distinction is reflected in our effective action by the sign of $c_{(4)}$; this coefficient is positive for the choice of coupling $f_1(\scalar)$ and negative for $f_2(\scalar)$.
Our simple energetics arguments reveal that above some critical electric charge (e.g., the green and blue curves with $c_{(2)}<0$), no spontaneously scalarized solutions exist, whereas below this value (e.g., the red and orange curves with $c_{(2)}>0$) spontaneously scalarized solutions may exist, but are unstable to scalar perturbations.
In the former case (the green and blue curves), there is no sign of a nonlinear saturation of the tachyonic instability of the EM solution; no stable equilibrium solutions seem to exist.
However, it is impossible to infer how the unstable EM solutions would evolve using our effective theory, since the assumption of time-reversal symmetry (or constant BH entropy) will likely break down.
Numerical-relativity simulations are needed to answer this question (or the construction of a more generic effective theory).

The importance of nonlinear interactions in stabilizing spontaneously scalarized solutions has been studied extensively in the context of ESTGB theories \cite{Blazquez-Salcedo:2018jnn, Minamitsuji:2018xde, Silva:2018qhn}.
For those theories, exponential couplings (equivalent to our $f_1$) or quartic couplings ($f \sim  - \phi^2 +\phi^4$) provide stable scalarized solutions, whereas with quadratic couplings ($f \sim  - \phi^2$), all scalarized solutions are unstable.
Interestingly, quadratic couplings predict stable scalarized solutions in EMS theories \cite{Myung:2019oua}, but our analysis suggests that stability is not guaranteed for generic couplings.
The stability analyses in these references involve studying linearized perturbations on scalarized backgrounds.
Though technically only valid near the critical point of the spontaneous scalarization phase transition and for small $q$, our approach offers a much easier alternative for estimating stability.
We find that our approach correctly reproduces the findings of these stability analyses for scalarized BHs in EMS theories~\cite{Myung:2018jvi,Myung:2019oua}.

\subsection{Critical exponents of phase transition in gravity}

The point-particle action~\eqref{SCO} also offers some insight into the critical behavior that arises near the onset of spontaneous scalarization.
For this discussion, we restrict our attention to the scalar coupling $f_1(\phi)$, for which spontaneously scalarized configurations are stable.
Considering the various coefficients $c_{\ldots}$ as functions of the electric charge $\e$ and entropy $\mathcal{S}$ of a BH and the overall coupling constant $\alpha$, the effective potential $V(\q)$ corresponds precisely to the standard Landau model of second-order phase transitions \cite{Landau:1937,*LandauTranslation}.
Compared to the archetypal example of ferromagnetism, the role of temperature $T$ is played by either $\e$ or $\alpha$.
This connection reveals that (i) spontaneous scalarization is a second-order phase transition and (ii) the critical exponents characterizing this phase transition match the universal values predicted by the Landau model.
Point (i) was already demonstrated for NSs in ST theories in Ref.~\cite{Sennett:2017lcx}, but point (ii) is new to this work; we elaborate on (ii) below.

Critical exponents dictate how a system behaves close to a critical point (e.g., the location of a second-order phase transition).
Such phenomena have first been discovered in GR in the context of critical collapse \cite{Choptuik:1992jv,Abrahams:1993wa,Evans:1994pj,Choptuik:1996yg}, but also appear in perturbations of extremal BHs \cite{Gralla:2017lto,Gralla:2018xzo}.
Applied to the current example of spontaneous scalarization, we study how the structure of the BH solutions varies as we approach the critical point at which spontaneous scalarization first occurs, parametrized by $\xi \rightarrow 0$ where $\xi$ could be either $\xi = (\alpha - \alpha_c) / \alpha_c$ at fixed $\e$ (identifying temperature as $T \sim - 1 / \alpha$) or $\xi = (\e - \e_c) / \e_c$ at fixed $\alpha$ (identifying $T \sim 1/\e$).
For example, the critical exponent $\beta$ of a Landau model is given by the scaling of the order parameter $\q \propto \xi^\beta$ as $\xi \to 0^+$.

The effective potential $V(\q)$ in Eq.~\eqref{eq:effEOM} depends on the properties of the BH solution; this dependence is suppressed in the notation used in the previous section, but here we explicitly restore it.
In particular, close to the critical point, the potential takes the form
\begin{align} \label{eq:EffPotentialTau}
V(\q;\xi) = \frac{c_{(2)}(\xi) }{2} \q^2 + \frac{c_{(4)} (\xi)}{4!} \q^4.
\end{align}
If $c_{(2)}(\xi) $ and $c_{(4)}(\xi) $ are analytic functions, they must take the following form near $\xi = 0$,
\begin{align}
c_{(2)}(\xi) = a \,\xi + \Order\left(\xi^2\right), \qquad
c_{(4)} (\xi)= b + \Order\left(\xi\right),
\end{align}
where $a$ and $b$ are positive constants.
Then, the minima of $V(\q;\xi)$ occur at
\begin{align}
\q=\pm \sqrt{\frac{-6c_{(2)}}{c_{(4)}}} = \pm \sqrt{\frac{-6a}{b}} \, \xi^{1/2},
\end{align}
thus, $q \propto \xi^{1/2}$ as $\xi \rightarrow 0^+$.
For this system, $q$ represents an order parameter, and thus the critical exponent $\beta$ is 1/2.

\begin{figure}
\includegraphics[width=\columnwidth]{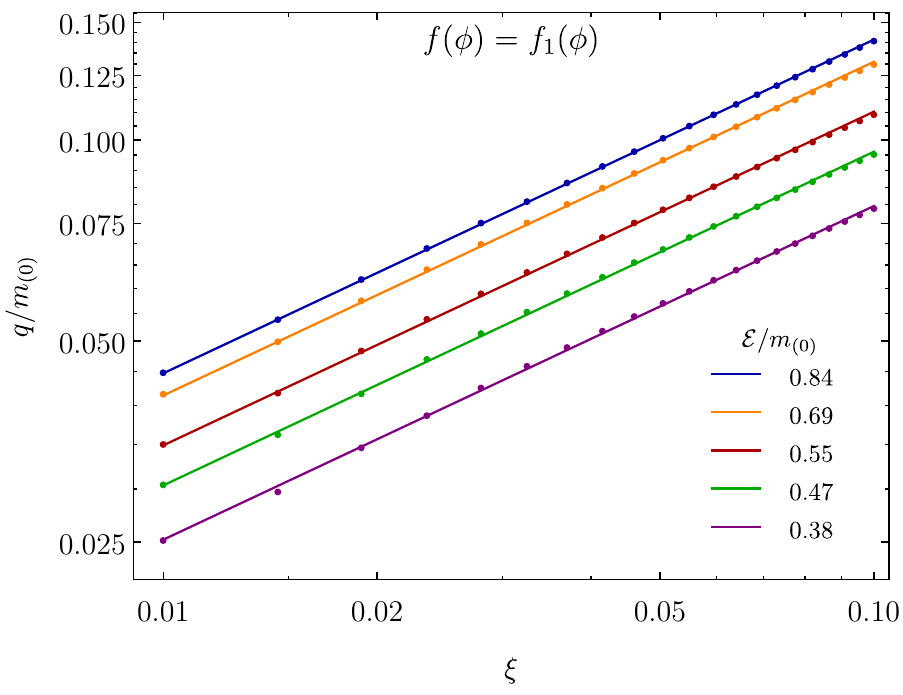}
\caption{\label{fig:expbeta}Scalar charge $\q$ as a function of $\xi\equiv (\alpha-\alpha_c)/\alpha_c$ for different electric charges with coupling $f_1(\phi)$~[Eq.~\eqref{eq:f1coupling}]. 
The numerical solutions are indicated with points, while the solid lines are best fits with slope 1/2.
We see that the solutions agree well with the expected scaling $\q \propto \xi^{1/2}$.
}
\end{figure}

We numerically confirm this claim by computing the scalar charge $\q$ of electrically charged BHs as a function of $\xi$ near the critical point.
Fixing the electric charge $\e$, we first determine the critical coupling $\alpha_\text{crit}$ at which $c_{(2)}$ vanishes.
We then compute $\q$ for couplings just below this value, i.e., for $\xi = (\alpha - \alpha_\text{crit})/\alpha_\text{crit} \gtrsim 0$.
The dependence of $\q$ on $\xi$ is depicted in Fig.~\ref{fig:expbeta}; we find that $\q \propto \xi^{1/2}$ agrees well with our numerical results.

Similarly, we compute the other standard critical exponents describing the phase transition.
In particular, both the analytic model~\eqref{eq:EffPotentialTau} and our numerical solutions indicate that $\gamma = 1$ where $\chi=d\q/d\scal0  \propto |\xi|^{-\gamma}$ for $\xi \rightarrow 0^{\pm}$, and $\delta = 3$ where $\q\propto \scal0^{1/\delta}$ at $\xi = 0$.
These findings are consistent with the Landau model for phase transitions.
It would be interesting to find a correspondence to a correlation length in the future, so that all standard critical exponents can be studied.
The introduction of a correlation length (becoming infinite as $c_{(2)} \rightarrow 0$) as another scale next to the size of the compact object could also allow for a more formalized power counting for the construction of the effective action close to the critical point.

\subsection{Dynamical scalarization}

We now employ our effective action to study the dynamical scalarization of binary systems in EMS theory, only considering the scalar coupling $f_1(\phi)$ except where noted.
For this purpose, we integrate out the remaining IR fields from the complete action (i.e., the field part, suitable gauge-fixing parts, and a copy of $S_\text{CO}^\text{crit}$ for each body).\footnote{Strictly speaking, the IR fields are split again into body-scale and radiation-scale parts, and we integrate out the body-scale fields~\cite{Goldberger:2004jt}. This would be necessary for a treatment of radiation from the binary using effective-field-theory methods~\cite{Goldberger:2009qd}.}
We employ a weak-field and slow-motion (i.e., post-Newtonian, PN) approximation.
These approximations are not independent here, since a wide separation of the binary (weak field) implies slow motion due to the third Kepler law for bound binaries.
The leading order in this approximation is just the Newtonian limit of the relativistic theory we are considering.
Therefore, the Lagrangian of the binary to leading order (LO) reads\footnote{We have also gauge-fixed the worldline parameters to the coordinate time $\sigma=t$ as usual in the PN approximation.}
\begin{equation}
  \begin{split}
  L^\text{LO} &= - \m_A \left[ 1 - \frac{\dot{\bm{y}}_A^2}{2} \right] - \m_B \left[ 1 - \frac{\dot{\bm{y}}_B^2}{2} \right]
  + \frac{c_{\dot{\q}^2,A}}{2} \dot{\q}_A^2 \nl
  + \frac{c_{\dot{\q}^2,B}}{2} \dot{\q}_B^2
  + \frac{\m_A \m_B}{r} + \frac{\e_A\e_B}{r} + \frac{\q_A \q_B}{r},
  \end{split}
\end{equation}
where $A$ and $B$ label the bodies, $r = |\bm{y}_A - \bm{y}_B|$ is their separation, and in this section the dot $\dot{~}$ indicates a derivative with respect to coordinate time.
We have suppressed the dependence of $m_A$ on $q_A$ for brevity, but recall that the ``free energy'' of each body takes the form
\begin{align}
m_A(q_A) &= m_{(0),A} + V(q_A) \nonumber\\
&= c_{(0),A} + \frac{c_{(2),A}}{2} q_A^2 + \frac{c_{(4),A}}{4!} q_A^4 .
\end{align}
Notice that $m_A(q_A)$ plays the role of the body's mass in the binary Lagrangian because (i) it couples to gravity like a mass, see Eq.~\eqref{SCO}, and (ii) it is independent of the fields, so that for the purpose of integrating out the fields it can be treated as a constant.
The Hamiltonian for the binary can be obtained via a Legendre transformation
\begin{equation}
\begin{split}
H^\text{LO} &= \m_A + \m_B
+ \frac{\bm{p}_A^2}{2 \m_A} + \frac{\bm{p}_B^2}{2\m_B}
+ \frac{p_{q,A}^2}{2 c_{\dot{\q}^2,A}} + \frac{p_{q,B}^2}{2 c_{\dot{\q}^2,B}} \nl
- \frac{\m_A \m_B}{r} + \frac{\e_A\e_B}{r} - \frac{\q_A \q_B}{r},
\end{split}
\end{equation}
with the pairs of canonical variables $(\bm{y}_{A/B}, \bm{p}_{A/B})$ and $(q_{A/B}, p_{q,A/B})$.

Following Ref.~\cite{Sennett:2017lcx}, let us now consider a special case that allows simple analytic solutions for the scalar charges of the bodies.
We henceforth assume that the scalar charges evolve adiabatically $p_{q,A/B} \approx 0$ and that the two bodies are identical, i.e., $\q\equiv \q_A = \q_B$, $c_{(2)}\equiv c_{(2),A} = c_{(2),B}$, etc.
The Hamiltonian in the center-of-mass system $\bm{p} \equiv \bm{p}_A = - \bm{p}_B$ now reads
\begin{equation}
H^\text{LO,adiab.} = 2 \m
+ \frac{\bm{p}^2}{\m}
- \frac{\m^2}{r} + \frac{\e^2}{r} - \frac{\q^2}{r}.
\end{equation}
Under these assumptions and recalling again that $m=m(q)$, the equation of motion for the scalar charge $\q$ is given by
\begin{equation}\label{qEOM}
0 \approx \dot{p}_{\q} = \frac{\partial H^\text{LO,adiab.}}{\partial \q} = 2 z \left(c_{(2)} \q + \frac{c_{(4)}}{6} \q^3\right) - \frac{2\q}{r},
\end{equation}
with the redshift
\begin{equation}
z \equiv 1-\frac{\mathbf{p}^2}{2 m^2}-\frac{m}{r} .
\end{equation}
For simplicity, we neglect relativistic corrections to the redshift from here onward, i.e., $z \approx 1$; restoring these corrections does not affect the qualitative behavior that we describe. 
Equation~\eqref{qEOM} has three solutions: an unscalarized solution with $\q=0$ and a two scalarized solutions with nonzero $q$ of opposite signs.
The condition for stability of these solutions is that they are located at a minimum of the energy of the binary,
\begin{equation}
0 \leq \frac{\partial^2H^\text{LO,adiab.}}{\partial \q^2} \approx 2c_{(2)} - \frac{2}{r} + c_{(4)}\q^2,
\end{equation}
which is violated for $\q=0$ when $1/r > c_{(2)}$.
Hence, the stable solutions are given by 
\begin{equation}\label{Qsoln}
\q = \left\{ \begin{tabular}{lll}
$0$ & \quad\text{for} & $1 / r \leq c_{(2)}$ \\
$\pm \sqrt{\dfrac{6}{c_{(4)}}} \sqrt{\dfrac{1}{r} - c_{(2)}}$ &
\quad\text{for} & $1 / r \geq c_{(2)}$
\end{tabular} \right. ,
\end{equation}
which contain a phase transition at $r = 1 / c_{(2)}$ corresponding to the spontaneous breaking of the $q\rightarrow -q$ symmetry of the effective action.
Recall that $c_{(2)} < 0$ corresponds to the case where each object is spontaneously scalarized, for which the bottom condition in Eq.~\eqref{Qsoln} always holds, and thus $q \neq 0$ over all separations.

\begin{figure}
	\includegraphics[width=\columnwidth]{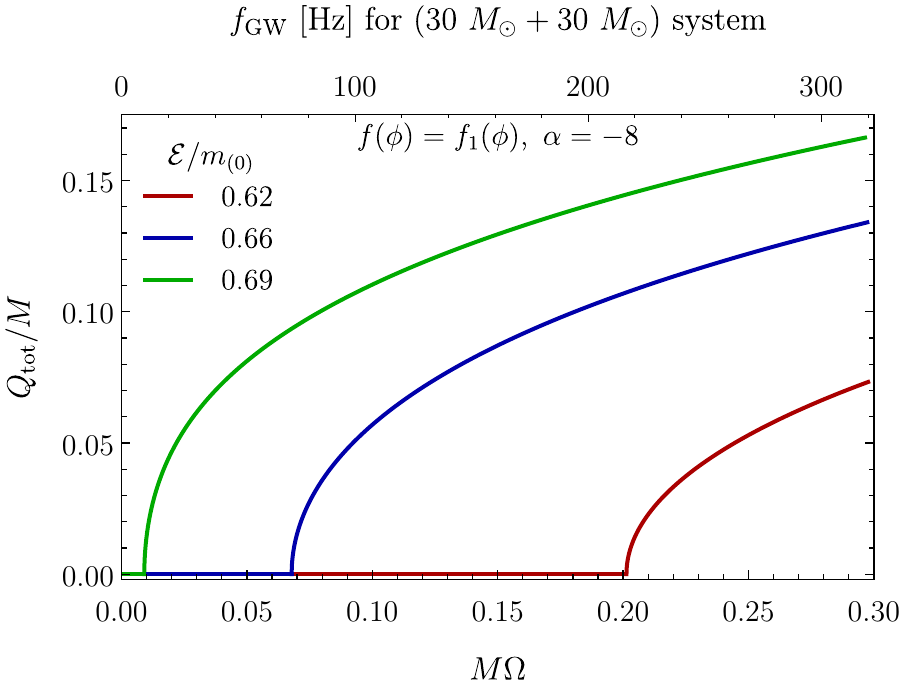}
	\caption{\label{fig:QOmega} Scalar charge $Q_\text{tot}$ of an equal mass binary as a function of orbital frequency $\Omega$ or GW frequency $f_\text{GW}=\Omega / \pi$ for coupling $f_1(\phi)$~[Eq.~\eqref{eq:f1coupling}]  with $\alpha=-8$.}
\end{figure}

Restricting our attention to circular orbits, we plot in  Fig.~\ref{fig:QOmega} the total scalar charge of the binary $\Q_\text{tot}\equiv \q_A+\q_B$ as a function of orbital frequency, given by Kepler's law as
\begin{equation}\label{eq:Kepler}
\Omega^2=\frac{1}{r^3}\left(\m_A+\m_B\right)\left(1+\frac{\q_A\q_B}{\m_A\m_B}-\frac{\e_A\e_B}{\m_A\m_B}\right).
\end{equation}
For simplicity, we only show the positive scalar charge branch of solutions.
The frequency is shown both as the dimensionless combination $M\Omega$ with $M\equiv \m_A^{(0)}+\m_B^{(0)}$ and as the equivalent GW frequency $f_\text{GW} = \Omega / \pi$ for a $(30 M_\odot + 30 M_\odot)$ binary system.
The plotted curves correspond to solutions with coupling constant $\alpha=-8$, and the colors correspond to different values of the electric charge.
The scalar charge vanishes below the onset of dynamical scalarization; the scalar charge grows abruptly at some critical frequency $\Omega_\text{scal}$ (as evidenced by kinks in the plotted curves) because dynamical scalarization is a second-order phase transition---see Ref.~\cite{Sennett:2017lcx} for a more detailed argument that dynamical scalarization is a phase transition.

\begin{figure}
	\includegraphics[width=\columnwidth]{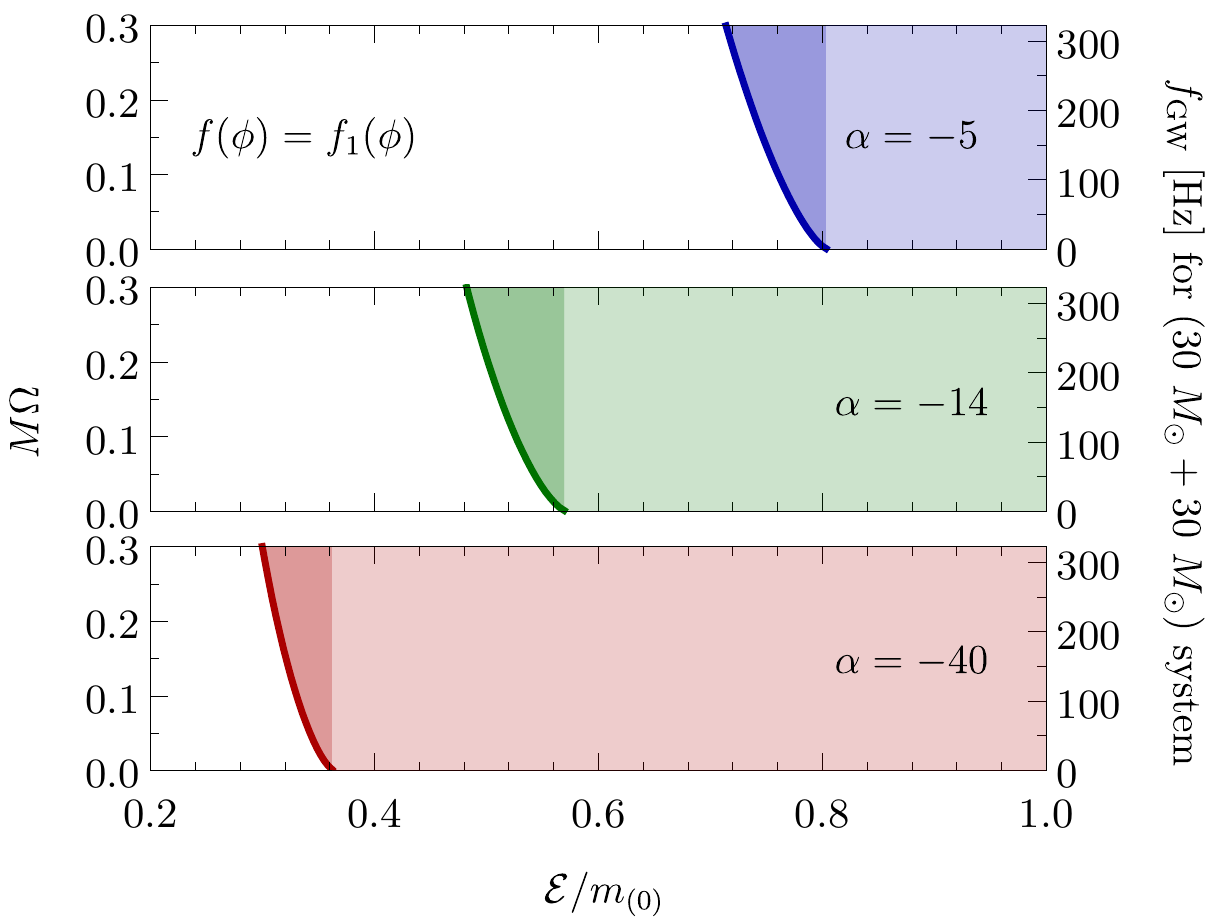}
	\caption{\label{fig:paramspace} Scalarization of binary systems with various electric charges for scalar coupling $f_1(\phi)$~[Eq.~\eqref{eq:f1coupling}] with different coupling strengths $\alpha$. Lightly and heavily shaded regions indicate spontaneously and dynamically scalarized configurations, respectively. The solid lines depict the onset of dynamical scalarization $\Omega_\text{scal}$ as a function of electric charge.}
\end{figure}

In Fig.~\ref{fig:paramspace}, we depict the scalarization of binary systems for various charge-to-mass ratios and couplings $\alpha$.
The solid lines indicate the critical frequency $\Omega_\text{scal}$ at which dynamical scalarization begins; the heavily shaded regions above these lines correspond to dynamically scalarized binaries after the onset of this transition.
The critical point $c_{(2)}=0$, corresponding to $\Omega_\text{scal} \rightarrow 0$, represents the division between binaries that dynamically scalarize and those whose component BHs (individually) spontaneously scalarize; we depict all spontaneously scalarized configurations with a lighter shading.
Thus, we see that our effective action, which was matched to isolated objects and models spontaneous scalarization, predicts dynamical scalarization as well.
Refined predictions can be obtained by perturbatively calculating the binary Lagrangian to higher PN orders and also the emitted radiation; both are possible using effective-field-theory techniques \cite{Goldberger:2004jt,Goldberger:2009qd,Kuntz:2019zef} or more traditional methods where extended bodies are represented by point particles, e.g., Refs.~\cite{Blanchet:2013haa,Bernard:2018hta}.

The same calculation can be repeated for binary systems with the choice $f(\scalar) = f_2(\scalar)$.
As before, the unscalarized $q=0$ solution is stable for above $r=1/c_{(2)}$.
However, because $c_{(4)} < 0$, no stable dynamically scalarized branch exists below that separation; instead, the system becomes ``dynamically'' unstable after this critical point.
The phase diagram for this choice of coupling takes the same form as Fig.~\ref{fig:paramspace}, but here the shaded regions correspond to scenarios in which no stable configuration exists.
At the onset of instability, the scalar radiation will likely grow rapidly and the GW frequency will decrease more rapidly compared to the EM case.
However, long-term predictions are not possible with our effective theory since the assumption of time-reversal symmetry likely breaks down, as for the unstable isolated BHs.

\section{Conclusions}
\label{sec:conclusions}

In this paper, we developed a simple energetic analysis of spontaneous and dynamical scalarization, based on a strong-field-agnostic effective-field-theory approach (extendable beyond the scalar-field case \cite{Ramazanoglu:2017xbl,Ramazanoglu:2018tig,Ramazanoglu:2018hwk} in the future).
We demonstrated our analysis for BHs with modified electrodynamics here, complementing the study of NS in ST gravity from Ref.~\cite{Sennett:2017lcx}. 
The theory-agnostic nature of our approach allowed us to draw general conclusions about scalarization
As an example, we found that dynamical scalarization generically occurs in theories that admit spontaneous scalarization.
Specific examples of theories for which our findings apply include those discussed in Refs.~\cite{Stefanov:2007eq,Doneva:2010ke,Ramazanoglu:2016kul,Doneva:2017bvd,Silva:2017uqg,Antoniou:2017acq,Antoniou:2017hxj,Doneva:2017duq,Ramazanoglu:2017xbl,Ramazanoglu:2019gbz,Ramazanoglu:2018tig,Ramazanoglu:2018hwk}.

The recent discovery of spontaneous scalarization in ESTGB theories~\cite{Doneva:2017bvd,Silva:2017uqg,Antoniou:2017acq,Antoniou:2017hxj,Doneva:2017duq} has sparked significant interest in this topic.
Our work predicts that dynamical scalarization can occur in binary systems in these theories and allows one to straightforwardly estimate the orbital frequency at which it occurs using only information derivable from isolated BH solutions.
Such information is valuable for guiding numerical-relativity simulations in these theories, for which there has been recent progress~\cite{Witek:2018dmd}.
Eventually higher PN orders in specific theories could be added to our model to derive more accurate predictions, and the framework could be extended to massive scalars or other types of new fields.

We demonstrated how scalarization, as an exemplary strong-gravity modification of GR, is parametrized by just a few constants in the effective action (which vanish in GR).
Hence, our effective action provides an ideal foundation for a strong-field-agnostic framework for testing dynamical scalarization.
Ultimately, one needs to incorporate self-consistently such effects into gravitational waveform models.
This undertaking will require the computation of dissipative effects (analogous to the standard PN treatment in GR) and the mode dynamics during scalarization, characterized (in part) by $c_{\dot{\q}^2}$.
The effective action can also, in principle, be extended to include other strong-field effects influencing the inspiral of a binary, e.g., phenomena like floating orbits \cite{Cardoso:2011xi,Endlich:2016jgc} or induced hair growth~\cite{Horbatsch:2011ye,Wong:2019yoc}.

Independent of GW tests of GR, further study of dynamical scalarization could offer some insight into the nonlinear behavior of merging binary NSs in GR.
As discussed in the Introduction, scalarization and tidal interactions enter models of the inspiral dynamics in a similar manner; in fact, the effective action treatment of dynamical tides (in GR) \cite{Steinhoff:2016rfi} is completely analogous to the approach adopted here for scalarization.
Unlike the case with tides, our model of scalarization includes nonlinear interactions via the $q^4$ term.
Nonlinear tidal effects could be relevant for GW observations of binary NSs~\cite{Reyes:2018bee,Weinberg:2018icl}, but are difficult to handle in GR.
Furthermore, mode instabilities also occur for NSs in GR \cite{Gaertig:2011bm,Andersson:2000mf}.
Dynamical scalarization can be used as a toy model for these types of effects, and further exploration of this non-GR phenomenon could improve gravitational waveform modeling in GR.

Our effective action approach allowed us to study the critical phenomena at the onset of scalarization; further study could also provide insight into critical phenomena in GR.
The critical exponents we obtained numerically agree with the analytic predictions from Landau's mean-field treatment of ferromagnetism \cite{Landau:1937,*LandauTranslation}.
Only missing here is a proper definition of correlation length, which we leave for future work.
In GR, the BH limit of compact objects has been suggested to play the role of a critical point and lead to the quasiuniversal relations for NS properties \cite{Yagi:2016bkt,Yagi:2013bca}.
These quasiuniversal relations are invaluable for GW science because they reduce the number of independent parameters needed to describe binary NSs, improving the statistical uncertainty of measurements.
In the BH limit (for nonrotating configurations), the leading tidal parameter vanishes \cite{Damour:2009vw,Binnington:2009bb,Kol:2011vg}, like $c_{(2)}$ at the critical point here.
Better theoretical understanding of the origin of these universal relations could help improve their accuracy; utilizing information from the critical phenomena at the BH limit is a compelling idea, and scalarization could serve as a toy model in that regard.

\acknowledgments
We thank Emanuele Berti, Caio Macedo, N\'{e}stor Ortiz, and Fethi Ramazanoğlu for useful discussions.

\appendix

\section{Effective action for compact objects away from critical point}
\label{sec:appendixA}

A crucial assumption made in the construction of the effective action~\eqref{SCO} is the near-criticality of the scalar mode $\q$.
The power counting used in the main text to formulate this relatively simple effective theory is not valid without this assumption.
For example, if $c_{(2)} > 0$ (i.e., the object does not spontaneously scalarize), then away from the critical point one finds $\q \sim \scals(y) \sim \Order(R/r)$ and must include terms like $[\scals(y)]^2$ to the action to work consistently at the given order in $R/r$.
For scalarized compact objects $c_{(2)} < 0$ far from the critical point, the scalar field $\scals(y)$ and mode $\q$ reach values too large for our polynomial expansion around zero to be valid.
If one expands the fields around their true (nonzero) equilibrium values instead, the effective action no longer respects the spontaneously broken scalar-inversion symmetry of the underlying theory.

In this appendix, we relax this assumption that the scalar mode $\q$ is nearly critical.
We construct an effective action valid in this broader context, and then show that the model~\eqref{SCO} is recovered as one approaches the critical point.
We still ignore derivative couplings involving $\partial_\mu \scals(y)$ since they belong to multipoles above the monopole ($\ell > 0$) and we are only interested in a monopolar mode $\q$.
The most generic effective action in the scalar-inversion-symmetric (unbroken) phase then reads
\begin{multline}\label{origaction}
  S_\text{CO}^\text{unbroken} = \int d\sig \bigg[
  \frac{c_{\dot{\q}^2}}{2} \dot{\q}^2
  - c_{(0,0)} - V - V^\scalar - V^{\q\scalar} \\
  + c_A A_\mu^\text{IR}(\y) \dot{y}^\mu
  \bigg] ,
\end{multline}
\begin{align}
V &= \frac{c_{(0,2)}}{2} \q^2 + \frac{c_{(0,4)}}{4!} \q^4 + \dots , \\
\begin{split}
V^\scalar &= \frac{c_{(2,0)}}{2} [\scals(y)]^2 + \frac{c_{(4,0)}}{4!} [\scals(y)]^4 \nl
  + \text{time derivatives} + \dots ,
\end{split}\\
\begin{split}
  V^{\q\scalar} &= - \scals(\y) \q
  + \frac{c_{(1,3)}}{3!} \scals(y) \q^3
  + \frac{c_{(2,2)}}{4} [\scals(y)]^2 \q^2 \nl
  + \frac{c_{(3,1)}}{3!} [\scals(y)]^3 \q
  + \text{time derivatives} + \dots ,
\end{split}
\end{align}
where the subscripts in $c_{(i,j)}$ indicate the powers of $\scals$ and $\q$, respectively; the coefficients $c_{(n)}$ in the main text correspond to $c_{(0,n)}$ in this notation.
All terms must be even polynomials in $\scals$, $\q$ due to scalar-inversion symmetry and must contain an even number of time derivatives due to time-reversal symmetry.
Higher time derivatives that would appear in $V$ can always be removed by appropriate redefinition of $\q$ \cite{Damour:1990jh}; we assume that such field redefinition has been done.
This allows for an interpretation of $V$ as an ordinary potential for the mode $\q$ in the absence of an external driving field $\scals$.

To fix all coefficients in the action, one needs to match against the exact solution for an isolated body in a generic time-dependent external scalar field.
In general, this is a complicated endeavor, and we do not attempt it here.\footnote{Note that our assumption of time-reversal symmetry needs to be imposed on the exact solution as well, so, in the case of a BH, one must impose somewhat unphysical (reflecting) boundary conditions at the horizon.}
Instead, we explore what information can be gleaned from the sequences of equilibrium solutions considered in the main text.
Using only this restricted class of solutions, we do not expect to find a unique match for all coefficients in the effective action above, but rather a series of relations relating them to the exact solutions.

Solutions for compact objects in equilibrium are manifestly time independent, and thus cannot inform the terms containing time derivatives in the effective action; we omit these terms from the action below for brevity.
We perform the same  procedure outlined in the main text  to match the IR fields of the effective theory~\eqref{origaction} to the IR projection of the UV solutions~\eqref{eq:UVsols} and find $c_A = \e$, $\scal0 = \scals(y)$, and
\begin{align}
  \Q(\scal0) &= - \frac{\partial V^{\q\scalar}(\q, \scals)}{\partial \scals}
              - \frac{d V^{\scalar}(\scals)}{d \scals} , \label{eq:Qmatch} \\
  \mn(\scal0) &= c_{(0,0)} + V + V^\scalar + V^{\q\scalar} . \label{eq:Mmatch}
\end{align}
Together with the equation of motion for $q$,
\begin{equation}
0 = \frac{d V(\q)}{d \q}
 + \frac{\partial V^{\q\scalar}(\q, \scals)}{\partial q} , \label{eq:qEOMapp}
\end{equation}
we see that
\begin{align}
\begin{split}
d \mn &= \left[ \frac{d V(\q)}{d \q}
  + \frac{\partial V^{\q\scalar}(\q, \scals)}{\partial q} \right] d \q \nl
  + \left[ \frac{\partial V^{\q\scalar}(\q, \scals)}{\partial \scals}
    + \frac{d V^{\scalar}(\scals)}{d \scals} \right] d \scals
\end{split}\\
  & = - \Q \, d \scal0 ,
\end{align}
in agreement with the first law of BH thermodynamics \cite{Gibbons:1996af, Cardenas:2017chu}.
We see that $\scal0$ and $\Q$ are conjugate variables, and therefore we can construct the ``gravitational free energy''  $\mathbb{M}(\Q)$ via a Legendre transformation of $\mn(\scal0)$,
\begin{align}
  \mathbb{M}(\Q) \equiv & \mn(\scal0) + \scal0 \Q , \label{eq:Mbb}
\end{align}
such that $\scal0 = d \mathbb{M} / d \Q$.
As in the main text, from a sequence of exact compact-object solutions in the full theory, one obtains $\mn(\scal0)$ and $Q(\scal0)$ and then can compute $\mathbb{M}(\Q)$ numerically from Eq.~\eqref{eq:Mbb}.

To aid comparison, we also define
\begin{align}
\mathbb{V}(\Q) \equiv & \mathbb{M}(\Q) - c_{(0,0)} , \label{eq:Vcal}
\end{align}
which represents the component of the ``free energy'' due to the scalar charge $Q$.
It admits an expansion around $Q=0$
\begin{equation}
  \mathbb{V}(\Q) = \frac{C_{(2)}}{2} \Q^2 + \frac{C_{(4)}}{4!} \Q^4 + \dots, \label{eq:Vcalexpand}
\end{equation}
whose coefficients can be extracted numerically.
Unlike $V$, this quantity does not correspond to the potential of any dynamical variable, but instead simply represents the energetics of a sequence of equilibrium solutions.
While these two quantities are not directly related in general, in the vicinity of the critical point it is possible to reconstruct the potential $V$ from the energetics $\mathbb{V}$ (as we found in the main text).
In the remainder of this appendix, we demonstrate this connection explicitly by expressing the $C_{(n)}$  in terms of the coefficients in the effective action $c_{(i,j)}$, and then take the limit that $\q$ becomes unstable $c_{(0,2)} \rightarrow 0$, i.e., approaches the critical point.

Working perturbatively in $\scal0=\scals$, we solve the equation of motion~\eqref{eq:qEOMapp} for $q$, relate this solution to $\Q$ and $\mn$ via Eqs.~\eqref{eq:Qmatch} and~\eqref{eq:Mmatch}, and then insert these solutions into Eq.~\eqref{eq:Vcal} and read off the coefficients $C_{(n)}$ from the expansion in Eq.~\eqref{eq:Vcalexpand}.
We find that
\begin{align}\label{eq:coeffrel}
  C_{(2)} = & \frac{c_{(0,2)}}{1 - c_{(0,2)} c_{(2,0)}} , \\
  C_{(4)} =& \frac{1}{(1 - c_{(0,2)} c_{(2,0)})^4}  \left[c_{(0,4)} + 4 c_{(0,2)} c_{(1,3)} \right. \nonumber \\
  & \left.+ 6 c_{(0,2)}^2 c_{(2,2)} + 4 c_{(0,2)}^3 c_{(3,1)} + c_{(0,2)}^4 c_{(4,0)}\right] . \nonumber
\end{align}
We see that, in general, an instability in the mode $q$ cannot be inferred directly from $\mathbb{V}(Q)$, i.e., $C_{(2)} < 0  \nRightarrow c_{(0,2)}< 0$ and  $ c_{(0,2)}< 0 \nRightarrow  C_{(2)} < 0$.
However, close to the critical point $c_{(0,2)} \approx 0$, we find
\begin{align}
  C_{(2)} &= c_{(0,2)} + \Order(c_{(0,2)}^2) , \\
  C_{(4)} &= c_{(0,4)} + \Order(c_{(0,2)}),
\end{align}
which confirms the link between the dynamical mode potential $V(q)$ and the energetics of equilibrium solutions $\mathbb{V}(Q)$, close to the critical point.

The relation between the mode $\q$ and the scalar charge $\Q$ along a sequence of equilibrium solutions reads
\begin{align}
  q = &\frac{Q}{1 - c_{(0,2)} c_{(2,0)}}
  + \frac{Q^3}{3! (1 - c_{(0,2)} c_{(2,0)})^4}
  \Big[ c_{(1,3)} \nonumber \\
  &+ c_{(2,0)} c_{(0,4)} + 3 c_{(0,2)} (c_{(2,2)} + c_{(2,0)} c_{(1,3)}) \nonumber \\
  &  + 3 c_{(0,2)}^2 (c_{(3,1)} + c_{(2,0)} c_{(2,2)}) \nonumber \\
   & + c_{(0,2)}^3 (c_{(4,0)} + c_{(2,0)} c_{(3,1)}) \Big] + \dots \label{eq:qQrel}
\end{align}
Note that Eq.~\eqref{eq:qQrel} does not lead to $\Q = \q$ as in the main text when $c_{(0,2)} = 0$.
But full agreement with the main text (ignoring higher orders in $c_{(0,2)}$ throughout) is achieved if we redefine
\begin{equation}
  \q \rightarrow q + \frac{q^3}{3!} \left[ c_{(1,3)} + c_{(2,0)} c_{(0,4)} \right] + \Order(c_{(0,2)}, Q^5) .
\end{equation}

\section{Numerical calculation of $V(q)$}
\label{sec:appendixB}

In this appendix, we detail the numerical calculation of equilibrium BH solutions in EMS theory used to construct $V(q)$ through the matching procedure discussed in the main text.
A Mathematica code that illustrates this calculation is provided as supplemental material \cite{EMScode}.
We consider the class of theories in Eq.~\eqref{EMSaction} and
restrict our attention to static, spherically symmetric configurations.
Starting with the ansatz for vector potential and metric
\begin{gather}
A= \lambda(r) \, dt,\\
\label{metricansatz}
ds^2= -N(r)e^{-2\delta(r)}dt^2 +\frac{dr^2}{N(r)} +r^2(d\theta^2 + \sin^2\theta d\varphi^2),
\end{gather}
where $N(r)\equiv 1-2m(r)/r$ and $m(r)$ is the Misner-Sharp mass (not to be confused with $m(q)$ introduced in the main text),
the field equations reduce to
\begin{subequations}
	\label{ODEsr}
	\begin{align}
        &\lambda'=- \frac{e^{-\delta}}{f(\scalar)} \frac{\e}{r^2},\\
	&m'=\frac{1}{2}r^2N\scalar'^2 + \frac{\e^2}{2 f(\scalar) r^2} ,
	\\
	&\delta' + r\scalar'^2 =0, \\
	&(e^{-\delta}r^2N\scalar')' = - e^{-\delta}\frac{f'(\scalar) \e^2}{2 (f(\scalar))^2 r^2},
	\end{align}
\end{subequations}
where ${}' = d/dr$ and $\e$ is the electric charge of the BH.
Here, the field equation for the electric potential $\lambda$ was already integrated once, introducing the electric charge $\e$ as an integration constant.
We impose the following boundary conditions at the horizon
\begin{align}\label{BChoriz}
m(r_H)=&\frac{r_H}{2}, \quad \delta(r_H)=\delta_H, \quad \scalar(r_H)=\phi_H, \nonumber\\
\scalar'(r_H)=& - \frac{f'(\scalar_H)}{2 f(\scalar_H)r_H}\left(\frac{\e^2 }{f(\scalar_H)r_H^2-\e^2}\right).
\end{align}
Note that $\delta_H$ represents a simple rescaling of the time coordinate, and thus can be chosen arbitrarily; we ultimately rescale $t$ such that $\delta(r=\infty) = 0$.
For computational simplicity, we rescale all dimensional quantities by the horizon radius $r_H$, i.e., $\tilde{r}\equiv r/r_H$, $\tilde{m}\equiv m/r_H$, and $\tilde{\e}=\e/r_H$, and then compactify the domain over which they are solved using the variable
\begin{equation}
\label{xr}
x\equiv \frac{r-r_H}{r+b r_H} = \frac{\tilde{r}-1}{\tilde{r}+b},
\end{equation}
where the constant $b$ is chosen to adequately resolve the solution.

As discussed in the main text, we consider sequences of BH solutions with fixed electric charge $\e$ and entropy $\s$ (or horizon area), which is equivalent to fixed $\tilde{\e}$ and $r_H$.
Fixing these two parameters, we generate a sequence of solutions by solving Eqs.~\eqref{ODEsr} and~\eqref{BChoriz} for several values of $\phi_H$.
We then extract the mass $\mn$, asymptotic field $\scal0$, and scalar charge $\Q$ from the asymptotic behavior of the solution
\begin{align}
\label{BCinfty}
m(r)\to \mn + \Order\left(\frac{1}{r}\right), \quad \scalar(r)\to \scalar_0 + \frac{Q}{r} + \Order\left(\frac{1}{r^2}\right),
\end{align}
allowing us to implicitly construct the functions $\mn(\scal0)$ and $\Q(\scal0)$ used in the main text to compute the effective potential $V(\q)$.

\bibliography{inspire}

\end{document}